\documentclass[conference]{IEEEtran}
\IEEEoverridecommandlockouts
\usepackage{cite}
\usepackage{amsmath,amssymb,amsfonts}
\usepackage{algorithmic}
\usepackage{graphicx}
\usepackage{textcomp}
\usepackage{xcolor}
\usepackage{url}
\usepackage{acronym}
\usepackage{hyperref}
\usepackage{subcaption}
\usepackage{float}

\makeatletter
\newcommand{\linebreakand}{%
  \end{@IEEEauthorhalign}
  \hfill\mbox{}\par
  \mbox{}\hfill\begin{@IEEEauthorhalign}
}
\makeatother

\makeatletter
\def\ps@IEEEtitlepagestyle{%
\def\@oddfoot{\mycopyrightnotice}%
\def\@evenfoot{}%
}
\def\mycopyrightnotice{%
\gdef\mycopyrightnotice{}
}

\begin{document}

\acrodef{ABM}{Agent-Based Model}
\acrodef{CA}{Cellular Automata}
\acrodef{EBM}{Equation-Based Model}
\acrodef{ODE}{Ordinary Differential Equation}
\acrodef{PDE}{Partial Differential Equation}
\acrodef{DES}{Discrete Events Simulation}
\acrodef{DEVS}{Discrete Events System Specification}

%%
%% The "title" command has an optional parameter,
%% allowing the author to define a "short title" to be used in page headers.
\title{Design Patterns for Multilevel Modeling and Simulation}

\author{ \IEEEauthorblockN{ Luca Serena\IEEEauthorrefmark{1}, Moreno
    Marzolla\IEEEauthorrefmark{1}\IEEEauthorrefmark{2}, Gabriele
    D'Angelo\IEEEauthorrefmark{1}\IEEEauthorrefmark{2}, Stefano
    Ferretti\IEEEauthorrefmark{3} }
  \IEEEauthorblockA{\IEEEauthorrefmark{1}Department of Computer
    Science and Engineering, University of Bologna, Italy}
  \IEEEauthorblockA{\IEEEauthorrefmark{2}Center for Inter-Department
    Industrial Research ICT, University of Bologna, Italy}
  \IEEEauthorblockA{\IEEEauthorrefmark{3}Department of Pure and
    Applied Sciences, University of Urbino Carlo Bo, Italy} {\tt\small
    \{luca.serena2,moreno.marzolla,g.dangelo\}@unibo.it,
    stefano.ferretti@uniurb.it} \thanks{\color{red}\bf This is the
    author’s version of the article: ``Luca Serena, Moreno Marzolla,
    Gabriele D'Angelo, Stefano Ferretti, Design Patterns for
    Multilevel Modeling and Simulation, proc. 2023 IEEE/ACM 27th
    International Symposium on Distributed Simulation and Real-Time
    Applications (DS-RT'23), Singapore, October 4—5, 2023, pp 48—55''.
    \textcopyright 2023 IEEE. Personal use of this material is permitted.
    Permission from IEEE must be obtained for all other uses, in any
    current or future media, including reprinting/republishing this
    material for advertising or promotional purposes, creating new
    collective works, for resale or redistribution to servers or
    lists, or reuse of any copyrighted component of this work in other
    works. The publisher version of this paper is available at
    \url{https://dx.doi.org/10.1109/DS-RT58998.2023.00015}.}  }
\maketitle

\begin{abstract}
Multilevel modeling and simulation (M\&S) is becoming increasingly
relevant due to the benefits that this methodology offers. Multilevel
models allow users to describe a system at multiple levels of
detail. From one side, this can make better use of computational
resources, since the more detailed and time-consuming models can be
executed only when/where required. From the other side, multilevel
models can be assembled from existing components, cutting down
development and verification/validation time.  A downside of
multilevel M\&S is that the development process becomes more complex
due to some recurrent issues caused by the very nature of multilevel
models: how to make sub-models interoperate, how to orchestrate
execution, how state variables are to be updated when changing scale,
and so on. In this paper, we address some of these issues by
presenting a set of design patterns that provide a systematic approach
for designing and implementing multilevel models. The proposed design
patterns cover multiple aspects, including how to represent different
levels of detail, how to combine incompatible models, how to exchange
data across models, and so on. Some of the patterns are derived from
the general software engineering literature, while others are specific
to the multilevel M\&S application area.
\end{abstract}

\begin{IEEEkeywords}
Multilevel modeling, Design patterns, Agent-based models, Multiscale simulation.
\end{IEEEkeywords}

%%%%%%%%%%%%%%%%%%%%%%
\section{Introduction}

Modeling and simulation are powerful tools that are used to study the
behavior of a complex system without the need of conducting
experiments on the ``real thing''. Over the years, many modeling
methodologies have been developed and are routinely used; different
research communities tend to have their own preferred set of tools,
e.g., \acfp{ABM} are frequently used in the social sciences, whereas
continuous (equation-based) models are frequently used to study the
diffusion of epidemics~\cite{serena2023}.

The most frequently used class of models are monolithic, meaning that
a single model takes care of the whole system (models that have some
internal structuring that merely derives from software engineering
good practices of decomposition are still considered
``monolithic''). Although monolithic models are appropriate for many
kinds of studies, they fall short when large, complex scenarios must
be investigated. Firstly, the possibility of combining existing models
can reduce development time and simplify verification and validation,
provided that the sub-models have already been properly
validated. Secondly, complex scenarios may be too large and/or too
complex to be evaluated at the maximum level of detail. These facts
have motivated the use of \emph{multilevel models}, sometimes referred
to as \emph{multilayer} or \emph{multi-resolution} models.

\begin{figure*}[ht]
\centering\includegraphics[width=\textwidth]{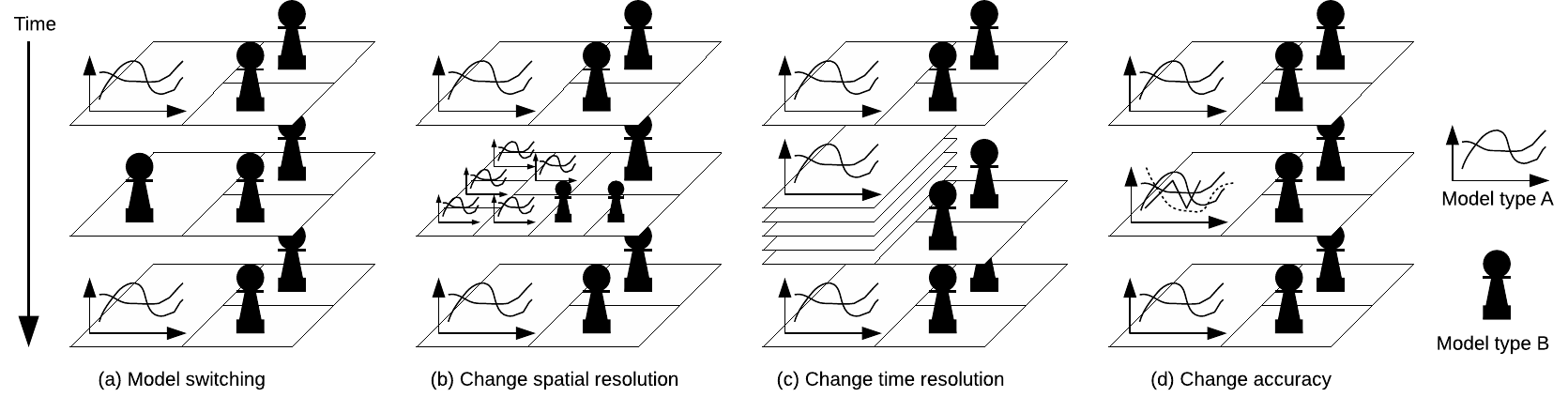}
\caption{The four main realizations of multi-level modeling.}\label{fig:multi-resolution-modeling}
\end{figure*}

Multilevel models employ multiple sub-models that may or may not be
active at the same time; sub-models can be based on different
paradigms (continuous, discrete, agent-based, stochastic, and so
forth), and may describe the system (or part thereof) at different
levels of detail. One key aspect of multilevel modeling is that the
decomposition into sub-models does not need to be static, as
Figure~\ref{fig:multi-resolution-modeling} depicts. In particular,
multilevel models are allowed to: (i)~switch any part of the system
under study to a different type of model (e.g., from continuous to
discrete models or back;
Figure~\ref{fig:multi-resolution-modeling}.a), (ii)~dynamically change
the spatial resolution of any part of the model
(Figure~\ref{fig:multi-resolution-modeling}.b), (iii)~dynamically
change the time resolution of any part of the model
(Figure~\ref{fig:multi-resolution-modeling}.c), or (iv)~dynamically
change the amount of state variables, i.e., the accuracy
(Figure~\ref{fig:multi-resolution-modeling}.d).

Multilevel techniques allow multiple level of details to be used
either for different portions of the system under study, or at
different points in simulated time. The goal is to make better use of
available computational resources: indeed, large and complex models
might not be executed efficiently if they are represented at the
maximum possible level of detail. Even when they could, they would
produce a large amount of data that, for the most part, would probably
be unimportant. Taking as an example a multilevel traffic model, we
might employ an equation-based representation of the aggregate flow of
vehicles, and switch to a more accurate agent-based model on areas
where interesting patterns emerge, such as road congestion; this
allows the simulation to ``zoom in'' and study in detail how
congestion develop and resolve. As soon as the traffic is back to
normal, the model can switch back to the faster, but less accurate,
equation-based model.

In the last decades many applications employing multilevel frameworks
have been used in several fields of study, e.g., human mobility,
traffic modeling, urban planning, social sciences, and others (see
Section~\ref{sec:background}). Despite the advantages outlined above,
multilevel modeling brings some issues that must be addressed: how can
sub-models interact? How can sub-models be scheduled for execution to
make efficient use of the available computational resources? How can
sub-models be orchestrated?

The analysis of the scientific literature reveals that there are some
recurrent solutions to the issues above. In this paper, we summarize
these solutions into a set of design patterns, with the aim of
contributing towards a wider adoption of multilevel M\&S. Some of the
design patterns come directly from the software engineering domain;
others are specific to the multilevel M\&S domain. Although reasonably
complete, due to space constraints the collection of patterns
described here is not meant to be exhaustive; however, we believe that
it covers the most important aspects.

This paper is structured as follows: in Section~\ref{sec:background}
we introduce the concepts of multilevel modeling and design
pattern. Then, from Section~\ref{sec:orchestration} to
Section~\ref{sec:multiscale} we describe the design patterns grouped
by function. Finally, some concluding remarks are discussed in
Section~\ref{sec:conclusion}.

%%%%%%%%%%%%%%%%%%%%%%%%%%%%%%%%%%%%%%%%%%
\section{Background}\label{sec:background}

Several modeling and simulation paradigms exist. They can be
classified according to at least two dimensions: (i)~how the state
space is represented, and (ii)~how time is represented.

In \emph{continuous-space models} the state space is represented by
continuous values (e.g.,~real numbers). Conversely, in
\emph{discrete-space} models the state space is described used
discrete values; a classical example is discrete~\acf{CA}, where each
cell can assume a finite set of values. In some cases, both continuous
and discrete variables are used to represent the state space; in this
case we have \emph{mixed} models.

For what concerns time management, we similarly have
\emph{continuous-time} and \emph{discrete-time}
models~\cite{law2015simulation}. In continuous-time models, the state
space changes continuously over time; these models usually are based
on sets of~\acp{ODE}. In discrete-time models, the state space is
updated only at specific points in time. Additionally, there are
\emph{time-stepped} models, where the time is divided into regular
discrete time frames, each one potentially corresponding to a defined
length of time. Finally, \emph{stochastic models} such as Monte Carlo
simulations do not rely on any representation of time, since they are
concerned with the computation of numerical results using stochastic
processes involving a large number of simulated experiments.

Different choices of space and time representation lead to different
types of models; to cite a few:
\begin{itemize}
\item \acp{EBM} are usually continuous-space, continuous-time and
  based on differential equations~\cite{van1998agent}; \acp{EBM} can
  usually be handled efficiently and are well suited for describing
  aggregate parameters over very large populations of entities.
\item \acp{ABM} are usually continuous-space, discrete-time and
  consist of a collection of autonomous, interacting computational
  objects (agents) that are situated in space and
  time~\cite{de2014agent}. \acp{ABM} are well suited for representing
  systems where interactions among agents, and between agents and the
  environment, are particularly important. However, \acp{ABM} can be
  computationally demanding, particularly if the number of agents is
  high.
\item \ac{CA} are usually discrete-space, time-stepped (although all
  possible variants exist); \ac{CA} represent the domain as a lattice
  of cells with simple rules to update the state of each cell based on
  the state of a subset of neighbors. Many classes of \ac{CA} can be
  evaluated quite efficiently through parallel computations.
\end{itemize}

In multilevel models, different paradigms can
coexist~\cite{ghosh1986concept}; sub-models might be either
semantically distinct (e.g.,~an urban simulation where different
components describe pedestrian mobility, traffic flows, air pollution,
land use and so forth), or describe the same item at different levels
of detail (e.g.,~traffic models that switch from~\acp{EBM} for
aggregate traffic flows and detailed~\acp{ABM} that describe
individual vehicles)~\cite{serena2022multilevel}.

The application spectrum of multilevel M\&S techniques is very
wide. In biology, chemistry or material science macroscale continuum
models can be used to simulate the behavior of fluids, solids, and
other materials, molecular simulators can study the dynamics and the
interactions of atoms and molecules, while an intermediate scale can
be used to simulate the behavior of larger groups of molecules, such
as polymers or proteins~\cite{dans2016multiscale,
  martins2010multiscale}.  In crowd and traffic simulation, usually
an~\ac{ABM} describes the behavior of individual pedestrians and
vehicles~\cite{nguyen2021overview}, while a macroscopic model deals
with equations that describe an aggregate high-level view of the
system~\cite{haman2017towards, cristiani2014multiscale}.  Finally, a
recurrent scheme to study the diffusion of epidemics is to have
within-host models that describe pathogen-host interactions, taking
into consideration immune system responses and the effect of
therapies, and between-host models that capture the dynamics of the
infection as it spreads from individual to
individual~\cite{mideo2008linking, almocera2018multiscale,
  qesmi2015immuno}.

Multilevel techniques simplify the development of complex models,
because they allow code reuse from existing sub-models, cutting down
development and validation times. Another key aspect is the
possibility of changing the level of detail or the type of paradigm at
run-time, to get a suitable trade-off between computational efficiency
of coarse-grained models and accuracy of fine-grained
representations. Indeed, it is often the case that only some critical
parts of a simulation are worth being represented at a high level of
accuracy, so using the maximum level of detail everywhere might be a
waste of time and resources.

However, multilevel M\&S techniques raise several issues, some of
which are concerned with software engineering aspects (e.g.,~how to
integrate components that were not necessarily created to work
together), while others are domain-specific (e.g.,~how to ensure
consistency among sub-models). Although the concrete solutions of
these issues are problem-specific, there are some recurrent patterns
that are frequently used in the literature.

In software engineering, \emph{design patterns} are standardized and
reusable solutions to recurrent software design problems that have
been proven to work effectively in practice. Design patterns are not
algorithmic solutions; rather, they are abstract descriptions of
solution schemes to classes of problems that must be instantiated to
each specific problem.

Conceptualized for the first time by Gamma et
al.~\cite{gamma1995design}, software design patterns have become an
important tool for helping developers to design high-quality,
maintainable, and efficient software systems.

\begin{figure*}
\centering\includegraphics[scale=.45]{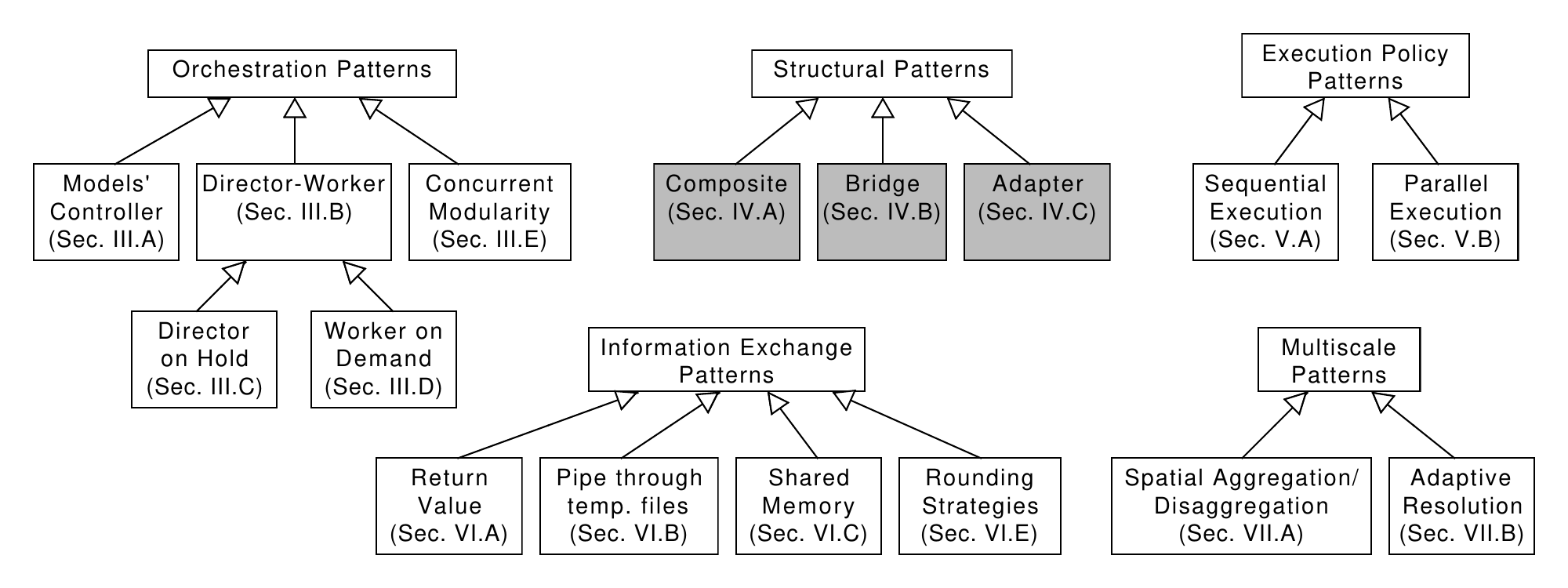}
\caption{Summary of the patterns described in this paper. Those originally described in~\cite{gamma1995design} are shaded.}\label{fig:list-of-patterns}
\end{figure*}

In this paper, we describe some design patterns that are found in
multilevel M\&S applications. The patterns are classified into five
categories, as illustrated in Figure~\ref{fig:list-of-patterns}:
\begin{itemize}
\item \emph{Orchestration patterns} deal with the flow of execution of
  sub-models (Section~\ref{sec:orchestration}).
\item \emph{Structural patterns} describe how sub-components can be
  aggregated into complex models (Section~\ref{sec:structural}). Note
  that these patterns are taken directly from~\cite{gamma1995design},
  since they are relevant for multilevel modeling besides general
  Object-Oriented programming.
\item \emph{Execution policy patterns} specify the mapping between
  components and execution units (Section~\ref{sec:execution-policy}).
\item \emph{Information exchange patterns} define how data can be
  transferred between models of different types, e.g.,~continuous and
  discrete-space models (Section~\ref{sec:information-exchange}).
\item \emph{Multiscale patterns} define how models employing different
  levels of details can be integrated (Section~\ref{sec:multiscale}).
\end{itemize}

%%%%%%%%%%%%%%%%%%%%%%%%%%%%%%%%%%%%%%%%%%%%%%%%%%%%%%%
\section{Orchestration patterns}\label{sec:orchestration}

Multilevel models involve the execution of multiple components, that
may be of different types (e.g.,~continuous and discrete models), or
of the same type using different parameters (e.g.,~different
time-steps). The components may be organized arbitrarily, i.e.,~not
necessarily in a strict hierarchy.

Orchestration patterns define how execution is passed from one
component to another. In the \emph{Model's Controller} pattern,
sub-models are executed by an external entity called
\emph{Controller}, who acts as an interface to the user. In the
\emph{Director-Worker} pattern, control is passed from the active
component to a different one. Finally, the \emph{Concurrent
  Modularity} pattern does not assume a strict hierarchical
structuring of sub-modules, and allows components to interact in a
peer-to-peer way.

The \emph{Director on Hold} and \emph{Worker on Demand} patterns are
possible realization of the Director-Worker paradigm. In the Director
on Hold realization, the Director is suspended until the worker(s)
terminates execution. The Worker on Demand pattern pre-allocates the
pool of workers in order to avoid the overhead of dynamically
creating/destroying them.

\begin{figure}[ht]
\centering\includegraphics[scale=.45]{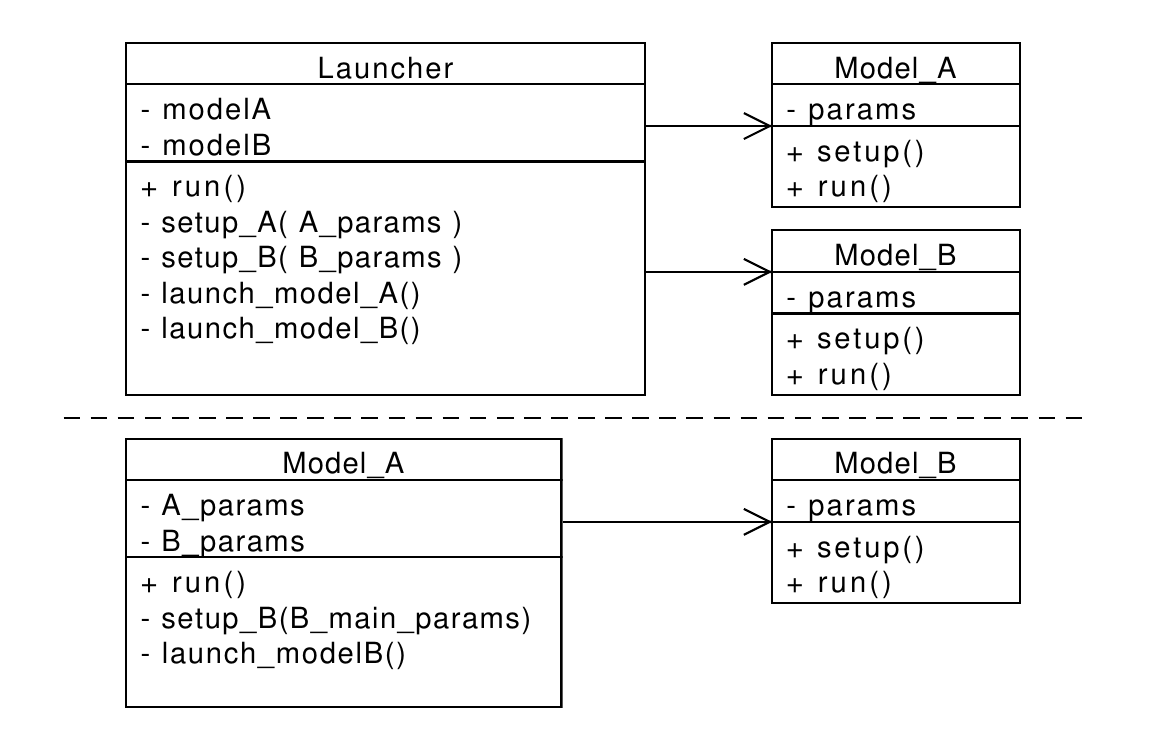}
\caption{Class diagrams of the Model's Controller (top) and Director-Worker pattern (bottom).}\label{fig:control-flow}
\end{figure}

\subsection{Models' Controller}

In this pattern there exists one entity, the Controller, that is in
charge of (i)~acting as the interface to the user or a higher-level
model; (ii)~scheduling the execution of the various sub-models;
(iii)~keeping a global state; (iv)~managing the exchange of
information among sub-models (see top of
Figure~\ref{fig:control-flow}).

The presence of a Controller has the advantage of centralizing the
scheduling and management logic, therefore allowing separation of
concerns between functionality and implementation. It also allows more
flexibility, as adding an additional sub-model is somewhat easier,
since only the Controller is involved. An obvious disadvantage is that
the Controller might become very complex if a large number of
incompatible sub-models are used.

The Controller pattern has been used in~\cite{xiong2013hybrid} to
investigate crowd evacuation using a multilevel model. The model
relies on a synchronization module to schedule execution of micro and
macro scales and manage the exchange of information.

\subsection{Director-Worker}

The Director-Worker pattern (bottom part of
Figure~\ref{fig:control-flow}) relies on a hierarchical structuring of
sub-models. Each sub-model can act as a worker with respect to its
parent module, and as a director with respect to children modules (if
any). Control is passed from a Director to a Worker. The Director
implements some of the functionalities of the Controller above;
however, unlike the Controller, a Director is itself a sub-model,
whereas the Controller is an external entity that is not part of the
model.  The Director-Worker pattern can be combined with the Composite
pattern (see Section~\ref{sec:structural}).

The Controller and Director-Worker patterns are not mutually
exclusive. For instance, in~\cite{d2018distributed} there is a
simulator at the top of the hierarchy that relies on a wrapper script
to manage the various instances of the underlying models. Thus, in
this case the wrapper script can be considered both as a Worker in a
Director-Worker scheme and as a Models' Controller of the lower-level
modules.

\subsection{Director on Hold}

The \emph{Director on Hold} pattern is the simplest realization of the
Director-Worker paradigm: the Director instantiates new Workers when
needed, and suspends itself while the Workers are active. At the end,
Workers are terminated and the Director resumes execution. Although
very simple, this strategy may incur a significant overhead if
creation/destruction of Workers is a costly procedure. Indeed,
while~\acp{EBM} might be cheaper to build and destroy, the same cannot
be said for~\acp{ABM}, as the creation of the agents and the storage
of their state is often a non-negligible activity. Furthermore, the
Director does not execute any computation while the Workers are
active, therefore reducing the level of concurrency that might be
allowed by the model.

\subsection{Worker on Demand}

\begin{figure}[ht]
\centering\includegraphics[width=\columnwidth]{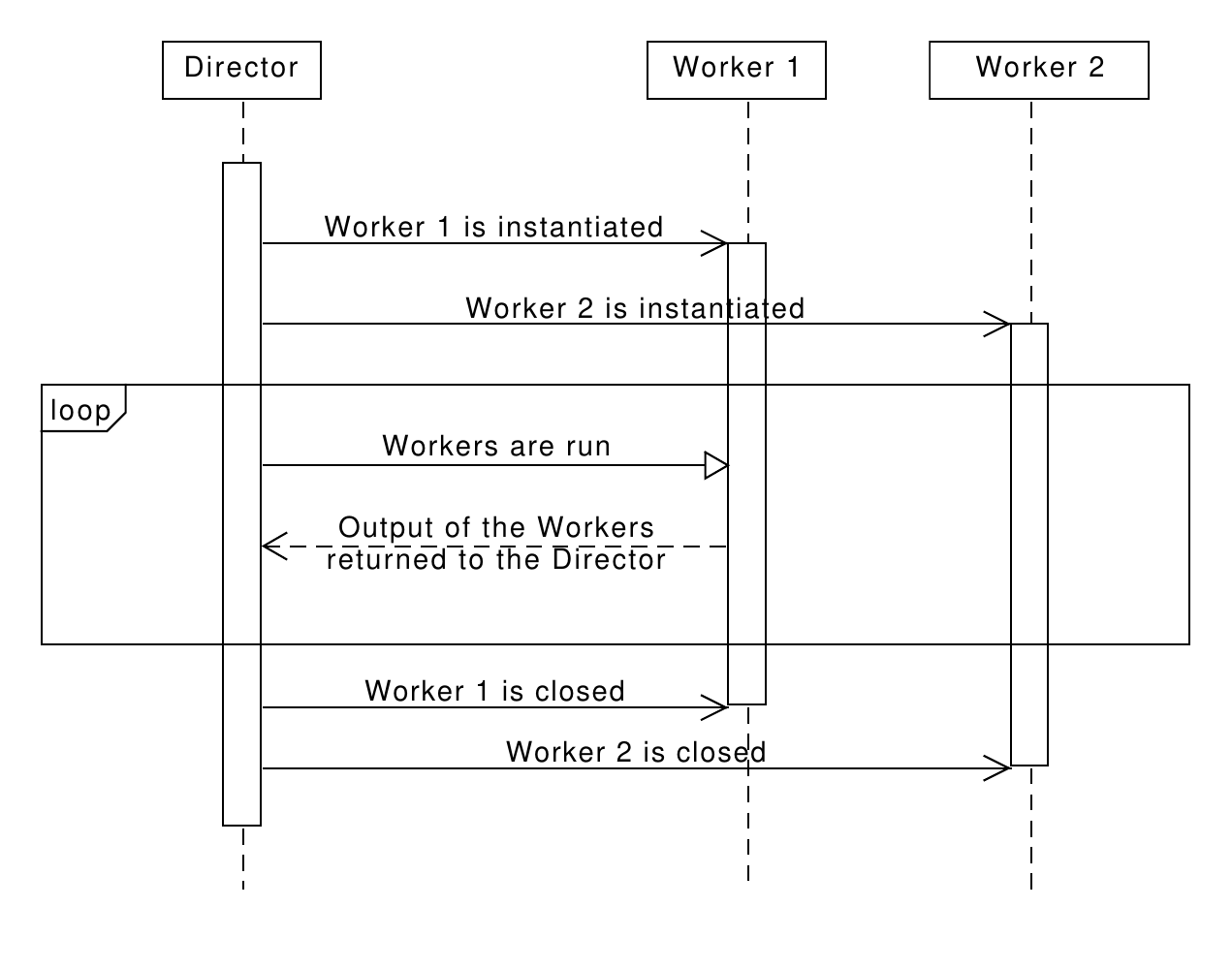}
\caption{Sequence diagram of Worker on Demand.}\label{WoD}
\end{figure}

This pattern addresses one of the limitations of the Director on Hold
pattern, namely, the overhead of creating/destroying Workers when
needed. In the Worker on Demand pattern, as shown in Figure~\ref{WoD},
all workers are created at the beginning of the execution and are kept
in stand-by; when one or more Workers are required, those in the pool
are dynamically assigned to complete some task.

The Worker on Demand pattern separates initialization of Workers from
the execution of tasks, with two main benefits: complex entities are
created only once (in case there is a large number of entities this
may save significant time during the life of the model), and the state
of the entities can be stored and retrieved for additional
examination. The drawback, however, is that the pool of Workers takes
up memory space even when inactive, making this strategy not
applicable in memory-constrained environments.

\subsection{Concurrent Modularity}

As already discussed, it is convenient to run multiple model instances
in parallel when possible. The \emph{Concurrent Modularity} pattern
allows concurrent execution of semantically different models, possibly
of different types. For example, a multilevel urban traffic model
might include sub-models for vehicle movements, pedestrian movements,
air pollution, and sound pollution. These sub-models might be executed
concurrently, provided that interactions are properly accounted for
(see Section~\ref{sec:information-exchange}).

The Concurrent Modularity pattern entails a thorough time management,
since different (sub-)models may use different time granularity and/or
different concepts of time; the latter happens, for example, when one
mixes continuous and time-stepped models.

The issue of time management can be addressed in different ways, such
as:
\begin{itemize}
\item A time translation mechanism, where the local time of a
  sub-model is translated into a global time understood by all other
  components.
\item Checkpointing, where sub-models proceed in lockstep and are
  synchronized periodically. A sub-model that reaches a checkpoint
  stops execution, and resumes when all other sub-models have also
  reached the same check-point.
\item Rollback mechanisms, where inconsistencies in state updates are
  detected and undone by rolling back the (virtual) simulation time to
  a previous time where a consistent state were
  computed~\cite{jefferson85}.
\end{itemize}

%%%%%%%%%%%%%%%%%%%%%%%%%%%%%%%%%%%%%%%%%%%%%%%%%%%
\section{Structural patterns}\label{sec:structural}

\emph{Structural patterns} describe how software elements can be
composed into larger structures while promoting flexibility and code
maintainability. The patterns described in this section are taken
from~\cite{gamma1995design}, where they have been initially proposed
in the context of software engineering.

\begin{figure}[ht]
\centering\includegraphics[scale=.45]{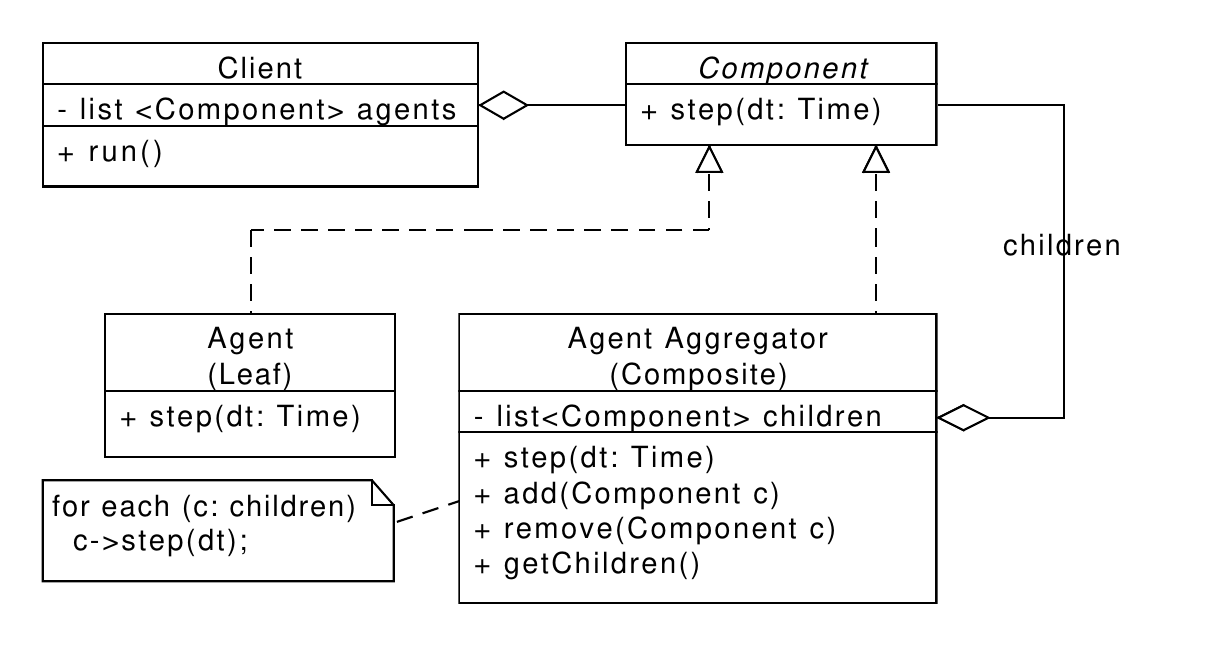}
\caption{UML class diagram of Composite pattern applied to a multiscale~\ac{ABM} scenario.}\label{fig:composite}
\end{figure}

\subsection{Composite}

The Composite pattern enables the hierarchical composition of objects
according to a tree-like structure, allowing atomic and composite
objects to be treated uniformly. This makes it easy to add new types
of objects to a system, as the interface for all components remains
the same. The pattern is composed of three main elements (see
Figure~\ref{fig:composite}):
\begin{itemize}
\item \emph{Component}, an interface that defines the common methods
  for both the Leaf and the Composite.
\item \emph{Leaf}, an end node of a tree structure.
\item \emph{Composite}, an internal node of a tree structure.
\end{itemize}

An application of this design pattern is in the context of
hierarchical~\acp{ABM}. Here, composite objects are actors at the
macro level that act as container of agents at the micro level. The
composite object may therefore represent a portion of a model that is
represented at a coarser level of detail; when more accuracy is
required, the composite executes the low-level agents that it
contains, which in turn might be composite objects and contain some
finer-grain sub-models.

This pattern could be applied in~\cite{mboup2017multi}, where the
authors studied the spread of black rats by means of commercial
transportation. In this work, the main building block of the simulator
is represented by the concept of \emph{World}, defined as a complete
and self-sufficient sub-model with its own places, agents, spatial
resolution and temporal scale. Worlds can possibly be nested,
representing part of another World at greater level of detail. The
Composite pattern could be then applied in order to provide a
high-level management of all the worlds in the system.

\subsection{Bridge}

The Bridge pattern allows developers to separate the abstraction
(interface) from the implementation~\cite{gamma1995design}.  The
Bridge pattern is composed of three main components:

\begin{itemize}
\item \emph{Abstraction}, which defines the high-level interface that
  clients will use.
\item \emph{Implementor}, which serves as an interface for describing
  the technical functionalities of the Abstraction.
\item \emph{Concrete Implementor}, which defines the concrete
  implementation of the Implementor.
\end{itemize}

\begin{figure}[ht]
\centering\includegraphics[scale=.45]{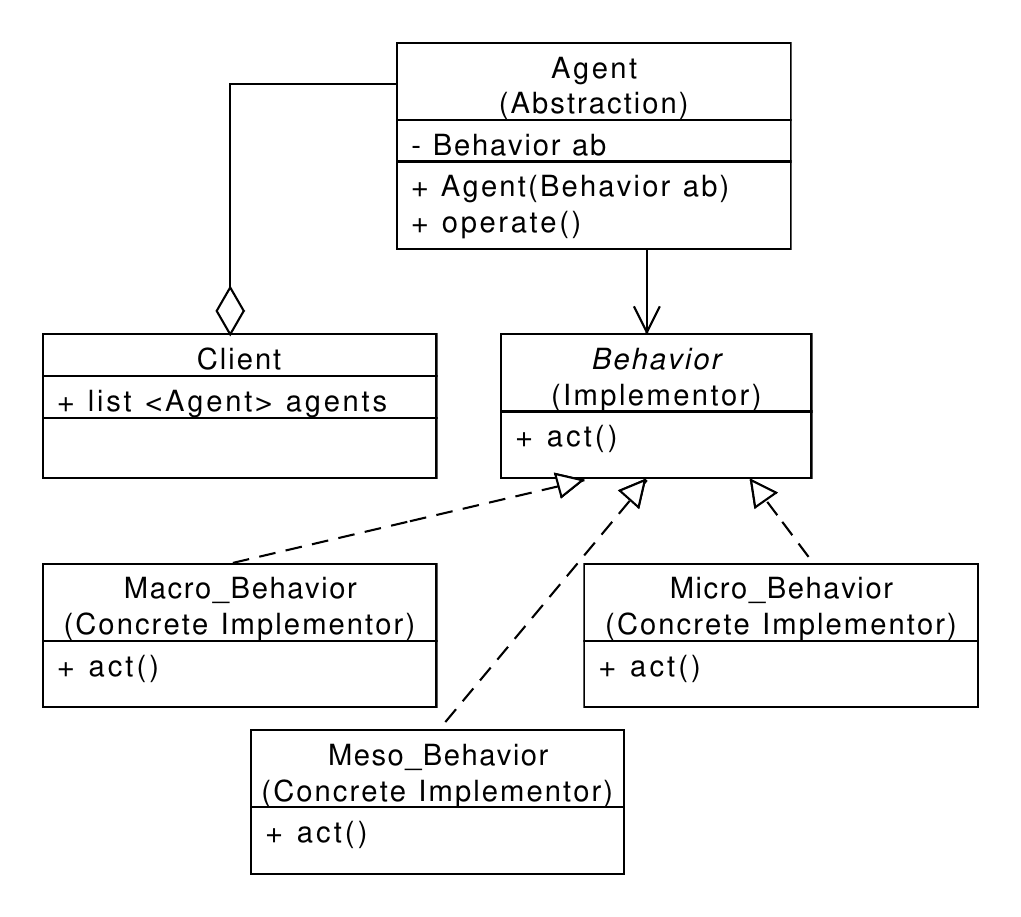}
\caption{Class diagram of the Bridge pattern applied to a multiscale \ac{ABM} scenario.}\label{bridge}
\end{figure}

Separating the implementation from the interface is one of the
cornerstones of Object-Oriented programming; among other things, it
allows different implementations of some abstract object to be
interchanged, even at run-time, without the need to modify clients
that are using the abstract objects.  The Bridge pattern can be
applied in~\acp{ABM}, enabling to separate the definition of the
agents from the code that defines the behavior of certain types of
simulated entity, as shown in Figure~\ref{bridge}.  For example, in a
simulation we could have different types of human agents (the
Abstraction), characterized by different types of behavior in response
to certain events.  Bridge can find an application also in the context
of multiscale modeling, for instance when individual and aggregate
agents coexist in the same simulated environment like
in~\cite{musse2001hierarchical}, where the movement of pedestrians is
simulated dealing with a hierarchy of crowds, groups, and individuals.
This pattern allows the developers to make changes to the behavior of
the agents without affecting how the Client interacts with them. Also,
by abstracting away the details of the implementation, the Abstraction
provides a simpler interface for the Client, improving the management
of the agents at a high-level.

\subsection{Adapter}

The \emph{Adapter} pattern allows two or more components with
incompatible interfaces to work together by creating an adapter that
converts one interface to another~\cite{gamma1995design}. The Adapter
pattern is composed of four components:
\begin{itemize}
\item \emph{Client}, the object that uses the Adapter to interact with
  the Adaptee.
\item \emph{Adaptee}, the object that needs to be adapted to work with
  the Client.
\item \emph{Target}, the object that the Client wants to use.
\item \emph{Adapter}, the object that acts as an intermediary between
  the Client and the Adaptee. The Adapter translates the interface of
  the Adaptee to the interface expected by the Client (i.e., the
  Target interface).
\end{itemize}

\begin{figure}[ht]
\centering\includegraphics[scale=.45]{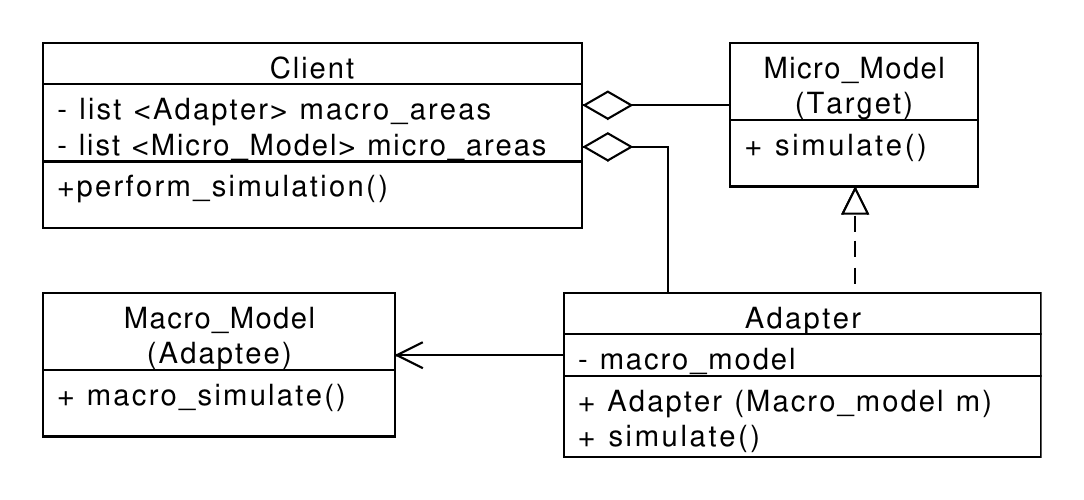}
\caption{UML class diagram of the Adapter pattern applied to a multiscale modeling scenario.}\label{fig:adapter}
\end{figure}

Since multilevel M\&S is based upon multiple cooperating components
(sub-models), the Adapter pattern is extremely useful when the
sub-models were not intended to work together. This is usually the
case when the multilevel model is built upon existing
components. Another use case for the Adapter pattern is shown in
Figure~\ref{fig:adapter}, and involves a multiscale traffic model.  In
these scenarios, the space is often split into micro zones where the
behavior of vehicles is modelled individually, and macro zones where
the traffic is represented in terms of aggregate metrics such as
density, average speed and traffic flow. Suppose that a traffic model
was initially developed using one level only, e.g., using only micro
zones. Then, the Adapter pattern can be employed to make the macro
zones (the Adaptees) look like micro zones from the point of view of
the coordinator (the Client).

%%%%%%%%%%%%%%%%%%%%%%%%%%%%%%%%%%%%%%%%%%%%%%%%%%%%%%%%%%%%%%%
\section{Execution policy patterns}\label{sec:execution-policy}

Execution policies refers to rules and guidelines that govern the
mapping of the model to the underlying execution unit(s). The choice
of execution policy depends on several factors, such as the size and
complexity of the model, the amount of computational resources
available, and the desired level of performance that should be
achieved.

\subsection{Sequential Execution}

In a sequential execution pattern, the model is executed on a single
execution unit, e.g., a single processor or processor core. This is
the most common approach, either because the vast majority of models
are inherently sequential, or because the existing implementations
are. Sequential execution is appropriate when the time required to
analyze a model is not the bottleneck.

\subsection{Parallel Execution}

Parallel execution involves splitting the model into smaller parts
that can be executed simultaneously on multiple execution units,
possibly over multiple interconnected machines. Parallel execution has
traditionally been applied to speed up the execution of monolithic
models, e.g., by processing events in parallel during discrete-event
simulations~\cite{fujimoto} or by employing parallel solvers for
analyzing large sets of differential equations.

In the context of multilevel M\&S, parallelization may be useful when
some form of domain partitioning is used to split the simulation space
into separate partitions, each one being modeled at a different level
of granularity and/or using different types of models. If the
partitions are independent, they can be evaluated in parallel,
although in practice the level of parallelism might be reduced because
neighboring partitions could need to exchange data periodically, or
some form of global consistency must be ensured.

%%%%%%%%%%%%%%%%%%%%%%%%%%%%%%%%%%%%%%%%%%%%%%%%%%%%%%%%%%%%%%%%%%%%%%%
\section{Information Exchange Patterns}\label{sec:information-exchange}

Information Exchange patterns define how data can be exchanged between
sub-models.  Different factors can affect the way in which the
information is transferred, such as the type of information (i.e.,
discrete vs continuous), or the relationship between the involved
components (i.e., hierarchical vs peer-to-peer).

\subsection{Return Value}

The \emph{Return Value} pattern is the most trivial way to exchange
data in a Director-Worker scenario. The Director calls The Worker that
returns back to the Director some result.  This strategy is very
simple but is difficult to implement if the Director is allowed to
execute concurrently with the Workers. In this scenario, the execution
of Workers is an asynchronous operation that may return well before
the Workers terminate. This problem can be addressed using the
\emph{futures} pattern used in concurrent programming~\cite{liskov88},
where the Worker returns a object representing the ``promise'' to
compute a result; the Director will block if it tries to read the
result when it has not been computed yet.

\subsection{Pipe through Temporary Files}

Another way to exchange information is through temporary files. This
provides the following advantages:

\begin{itemize}
\item \emph{Flexibility}: it is easy to add fields within the list of
  values to be returned.
\item \emph{Versatility}: it can be used also for flat models, since
  the consumers only need to know the location of the data file(s),
  and possibly when new data are available.
\item \emph{Data organization}: developers may choose the most
  appropriate data representation (e.g., JSON, XML, ...) depending on
  the type of information that needs to be stored and the requirements
  of the producers and consumers.
\end{itemize}
However, also known limitations must be considered:
\begin{itemize}
\item File operations (creation, read, write) entails some overhead,
  which can be relevant to a greater or lesser extent depending on the
  number of operations.
\item If the application aborts, the user must ensure that all
  temporary files are deleted before the simulator is launched again,
  to avoid pollution of a new execution with stale information.
\item Some additional mechanism must be put in place to inform the
  consumers when new data is available.
\end{itemize}

\subsection{Shared Memory}

In the \emph{Shared Memory} pattern, data is stored in some memory
region that is make accessible by all sub-models. Although this is
somewhat more efficient than using temporary files, shared memory
needs to be emulated on distributed-memory architectures. Furthermore,
if shared memory is used for two-way communication, special care
should be taken to avoid read-write conflicts.

\subsection{Rounding Strategies}

Exchanging data between sub-models that are base on continuous and
discrete state representations is a common problem in multilevel
modeling. In such scenarios, rounding strategies must be carefully
defined, in order to ensure global consistency properties such as
conservation of some model-specific entities.

As a practical example, let us consider a multilevel epidemic model
where a set of~\acp{ODE} is used to predict the diffusion of a
contagious disease through a population. To better study the contagion
in critical zones such as schools, hospitals, or crowded areas, the
model might delegate some regions to more detailed~\acp{ABM}. In this
scenario, it is essential that the total population is conserved,
i.e., assuming that the system is closed, the sum of susceptible,
infected, recovered and removed individuals must remain the same as
time progresses. However, \ac{ODE}-based models are continuous
while~\acp{ABM} are discrete, so we need to ensure that the values of
state variables -- in this case, the number of individuals in each of
the four classes above -- are stationary.

Several rounding strategies have been proposed in the literature: for
example, the model could keep track of rounding errors and add/remove
individuals in the~\acp{ABM} when needed.

%%%%%%%%%%%%%%%%%%%%%%%%%%%%%%%%%%%%%%%%%%%%%%%%%%%
\section{Multiscale patterns}\label{sec:multiscale}

Multiscale patterns deal with the issues related to representing the
same sub-model at different levels of detail. The use of multiscale
methodologies is motivated by the need to choose an optimal trade-off
between execution time and precision of the results.

\subsection{Spatial Aggregation-disaggregation}

A recurrent scheme of multiscale model is to have a microscopic level
where the behavior of the involved entities is described at a high
level of details, and a macroscopic level that deals with aggregate
metrics. The two levels might employ different types of models (e.g.,
\ac{ABM} for the micro level, and equation-based model for the macro
level), or the same type of model with different parameters (e.g., an
equation-based model for both levels using a finer time/space
subdivision to increase accuracy). The micro level is used when/where
``interesting'' phenomena emerge; for example, in a large urban
traffic model, the micro level would be used to focus on traffic jams
or transient congestion zones, in order to study how these pattern
form.

Aggregation and disaggregation is a modeling pattern that involves
collapsing a large number of entities at the micro level to build a
single entity at the macro level (aggregation), and the opposite act
of creating multiple entities at the micro level to represent a single
entity of the macro level (disaggregation). Therefore, the
Aggregation/Disaggregation design pattern establishes the rules by
which it is possible to switch between two levels of detail. This
pattern is often used in the context of~\ac{ABM}, although in
principle it can be applied to other types of models as well.

Multiple realizations of this design patterns have been described in
the scientific literature~\cite{mathieu2018multi}:
\begin{itemize}
\item \emph{Zoom} pattern, where the micro entities are destroyed when
  transitioning into the macro zones, and their information is lost.
\item \emph{Puppeteer} pattern, where the micro entities are not
  destroyed but frozen and temporarily controlled by the macro
  agents. Micro entities are still able to update their internal state
  according to their own dynamics, but cannot autonomously perform
  actions, which are delegated to the macro model.
\item \emph{View} pattern, where the state of micro entities is
  computed to reflect the state of the macro entities they emanate
  from.
\item \emph{Cohabitation} pattern, where the interactions between
  micro and macro entities are bidirectional, so they influence each
  other.
\end{itemize}

\subsection{Adaptive Resolution}

In many multiscale frameworks, the level of detail of a sub-model is
defined by its spatial or time resolution. In these scenarios it is
necessary to specify the conditions that trigger a change of
resolution. The \emph{Adaptive Resolution} pattern involves the
definition of these conditions, that is necessarily
model-dependent. As an example, in~\cite{bobashev2007hybrid} the
authors propose an adaptive multiscale infection propagation model
that combines the accuracy of~\acp{ABM} with the computational
efficiency of equations-based simulations. The model starts with the
agent-based paradigm in order to thoroughly represent the initial
dynamics of the diffusion of the pathogen, and then it switches to an
equation-based methodology after a certain threshold of infected
individuals is reached, so as to support a population-averaged
approach.

%%%%%%%%%%%%%%%%%%%%%%%%%%%%%%%%%%%%%%%%%%%
\section{Conclusions}\label{sec:conclusion}

In this paper we described a set of design patterns that can be used
to address some of the issues encountered in the development of
multilevel models.  The patterns are divided into five
categories. Orchestration Patterns are strategies to organize the
various building blocks; Structural Patterns are design solutions for
developing software systems with a hierarchical structure; Execution
Policy patterns provide a rationale for executing the building block
of a complex models; Multiscale patterns are solutions for
representing a system with multiple scales of detail; finally,
Information Exchange patterns describe how data can be exchanged among
the sub-models.

The novelty of this proposal is to bring the methodological
contribution provided by design patterns into the context of
multilevel modeling and simulation. In fact, ad-hoc answers for
recurrent modeling issues were missing from the state of art, as most
of the effort in multilevel M\&S has traditionally been devoted --
with some exceptions -- to application development rather than
methodological studies.

The list of patterns described here in not exhaustive; we are
currently working towards extending our collection by leveraging a
recent review of the state of the art~\cite{serena2023}.

\section*{Acknowledgements}

Moreno Marzolla was partially supported by the Istituto Nazionale di
Alta Matematica ``Francesco Severi'' -- Gruppo Nazionale per il
Calcolo Scientifico (INdAM-GNCS) and by the ICSC National Research
Centre for High Performance Computing, Big Data and Quantum Computing
within the NextGenerationEU program.

% Generated by IEEEtran.bst, version: 1.14 (2015/08/26)

\end{document}